\documentclass{article}
\pdfoutput=1

\usepackage{arxiv}

\usepackage[utf8]{inputenc} 
\usepackage[T1]{fontenc}    
\usepackage{hyperref}       
\usepackage{url}            
\usepackage{booktabs}       
\usepackage{amsfonts}       
\usepackage{nicefrac}       
\usepackage{microtype}      
\usepackage{lipsum}
\usepackage{booktabs,caption}
\usepackage[flushleft]{threeparttable}
\usepackage{longtable}
\usepackage{pdflscape}

\title{The Role of Individual User Differences in Interpretable and Explainable Machine Learning Systems}

\author{
  Lydia P. Gleaves\thanks{This work was sponsored by the Defense Advanced Research Projects Agency under Air Force Contract FA8750-19-C-1522. Opinions, interpretations, conclusions, and recommendations are those of the authors and are not necessarily endorsed by the United States Government. Approved for Public Release, Distribution Unlimited.} \\
  Department of Engineering Management and Systems Engineering\\
  The George Washington University\\
  Washington, DC 20052 \\
  \texttt{lpgleaves@gwu.edu} \\
   \AND
  Reva Schwartz \\
  Parenthetic, LLC\\
  Alexandria, VA \\
  \texttt{rschwartz@parenthetic.io} \\
     \AND
 David A. Broniatowski \\
 Department of Engineering Management and Systems Engineering\\
 The George Washington University\\
 Washington, DC 20052 \\
 \texttt{broniatowski@gwu.edu} \\
}

\begin{document}
\maketitle

\begin{abstract}
There is increased interest in assisting non-expert audiences to effectively interact with machine learning (ML) tools and understand the complex output such systems produce. Here, we describe user experiments designed to study how individual skills and personality traits predict interpretability, explainability, and knowledge discovery from ML generated model output. Our work relies on Fuzzy Trace Theory, a leading theory of how humans process numerical stimuli, to examine how different end users will interpret the output they receive while interacting with the ML system. While our sample was small, we found that interpretability -- being able to make sense of system output -- and explainability -- understanding how that output was generated -- were distinct aspects of user experience. Additionally, subjects were more able to interpret model output if they possessed individual traits that promote metacognitive monitoring and editing, associated with more detailed, verbatim, processing of ML output. Finally, subjects who are more familiar with ML systems felt better supported by them and more able to discover new patterns in data; however, this did not necessarily translate to meaningful insights. Our work motivates the design of systems that explicitly take users' mental representations into account during the design process to more effectively support end user requirements. 
\end{abstract}


\section{Introduction}
With the proliferation of machine learning (ML) applications into workplaces and everyday life, more people are relying on the output and decisions from complex ML models. This complexity has made it increasingly difficult for non-expert audiences to interact with and understand the output of automated technology. There is a widespread notion that increased transparency and the use of techniques that inform audiences \emph{how} systems work will lead to improved comprehension and trustworthiness (see \cite{adadi2018peeking} for a review). Yet, to date there is little empirical evidence to support that notion. Interpretability is especially recognized as an important feature of machine learning (ML) systems \cite{doshi2017towards}. Predictions are becoming increasingly accurate through the use of “black box” models, such as deep neural networks, which are nevertheless too complex to easily comprehend. Furthermore, these algorithms are increasingly used to make consequential societal decisions. 

Nevertheless, relatively little work has focused on how model output is actually processed and \textit{interpreted} by human evaluators. In this paper, we draw upon Fuzzy Trace Theory (FTT) \cite{reyna2012new} -- an empirically-validated theory of how humans interpret numerical stimuli -- to make predictions regarding how model interpretability varies between different human subjects (see \cite{martidoes} for an example of this variation among NASA engineers). According to FTT, model output is encoded into human memory as \emph{several simultaneous mental representations that vary in precision}. The most precise of these representations is referred to as the verbatim representation. For example, a model may generate an F1-score of 0.66. The verbatim representation of this error is simply “0.66”. However, this number can be difficult to interpret without context, such as the number of classes, and the balance between classes, in the model's training set. Additionally, the historical difficulty of the task matters, as do the consequences of false positives and false negatives for decision makers. Finally, users may be unfamiliar with how to interpret output scores, impeding their ability to connect these scores to their initial question, or to understand the underlying uncertainty, accuracy, or validity of the associated metric. Beyond the verbatim representation, humans encode several \textit{gist} representations of the model output which differ in terms of their levels of precision and reliance on context. For example, if the model is going to be used to make a prediction, the simplest categorical gist may distinguish between models that are “predictive” or “not predictive” relative to the state of the art. Even among two predictive models, a more precise ordinal gist may be used to distinguish between  a “better” or “worse” model fit depending on the context. Thus, gist representations incorporate background knowledge regarding how the model is to be used in practice -- its functional purpose. Gists may also be distinguished from one another by level of abstraction \cite{fukukura2013psychological, reyna2020abstraction}. For example, the gist of a model output -- its functional requirements or \textit{why} it is being used -- is quite distinct from the gist of its implementation details -- such as \textit{how} a given result is derived (see also \cite{rasmussen1985role}). In short, a user's ability to interpret, or \textit{make sense}, of a model's output, should be encoded distinctly from their ability to \textit{explain how} the model achieved its result. This motivates our first hypothesis:

\textbf{Hypothesis 1:} Mental representations for interpretability and explainability are  encoded distinctly and in parallel 

Importantly, individuals differ from one another in the degree to which they rely on more or less precise mental representations when making sense of stimuli. According to FTT, specific skills and personality traits are expected to predict reliance on more precise representation for mathematical tasks. Specifically, individuals who are more \textit{numerate} -- i.e., possess more facility with numbers -- have the \textit{ability} to operate on a detailed representation. Beyond numeracy, individuals with a higher \textit{Need for Cognition} \cite{cacioppo1984efficient}, possess the \textit{willingness} to rely on the verbatim representation when evaluating model output.

\textbf{Hypothesis 2:} Interpretability is significantly associated with data science expertise, numeracy, and need for cognition 

\section{Methods}

\subsection{Study Setting}
We conducted this study during a design event in which teams of software developers designed systems to automatically combine machine learning models (e.g., different regression and classification systems) into integrated pipelines. Four automated systems were evaluated in this work, each created by an independent team as part of the design event. Each system enabled a user to import a dataset, explore or interact with it, build a model, and reach some final decision.

The purpose of these systems was to solve problems specified by two different groups of end users: 1) subject matter experts (SMEs), and 2) data scientists. Each team was tasked with developing a system front-end to elicit input from end users, and then use that input to automatically construct a pipeline of machine learning models to solve problems specified by those end users. End users then evaluated the output according to its utility along several dimensions. Specifically, end users were given the opportunity to provide feedback to developers of these automated systems by ranking models according to their perceived utility. Thus, the task for developers was to automatically generate a set of models and propose them to the end user for downstream selection who would, in turn, select their preferred model. End users were then given the opportunity to visualize the system generated models, more clearly specify their problems, and edit the system proposed models. 

\subsection{Subject Recruitment}
The purpose of the study was to evaluate the functionality and usability of these prototype systems. We therefore recruited a convenience sample of subjects using an online sign-up page at http://parenthetic.io/studies. Subjects were recruited based on their stated interest and experience in specific domains. Specifically, we recruited subjects using snowball sampling from designated “gatekeepers” and supplemented this subject pool with a larger base of interested users that indicated relevant experience. Subjects from this user base received additional screening questionnaires to ensure they represented the targeted end user population.  

\subsection{Individual Traits Questionnaire}
All subjects completed a questionnaire containing several validated psychometric instruments designed to measure skills and personality traits that, according to Fuzzy-Trace Theory, should be associated with gist extraction. These traits include two measures of preference for effortful thinking: Need for Cognition \cite{cacioppo1984efficient} and the Cognitive Reflection Test \cite{frederick2005cognitive} and three measures of mathematical ability (numeracy) \cite{fagerlin2007measuring,cokely2012measuring,lipkus2001general}.  Additionally, we included a short form of the “Big Five” personality inventory \cite{soto2017short} to assess the effects of other commonly measured individual differences. Finally, subjects assessed their own data science expertise on a 6-point Likert scale ranging from ``I have no experience in data analysis'' (coded as 0) to ``I have extensive coursework and professional experience in data modeling and/or engineering'' (coded as 6; see Supplementary Materials).

All subjects completed the composite questionnaire remotely using Qualtrics survey software, with each instrument presented in a random order and items within each instrument presented in random order.  Responses for items in the Cognitive Reflection Test, Lipkus Numeracy Test, and Berlin Numeracy Test were marked correct or incorrect, and users were scored using the percent of items marked correct. The remaining inventories were scored according to their source.

\begin{table}
\centering 
\caption{Instruments used for composite questionnaire.}
\label{tab:booktabs}
\begin{tabular}{p{5cm}|p{5cm}|p{5cm}}
\toprule
\textbf{Instrument} & \textbf{Trait Measured} & \textbf{Sample Item} \\ [1ex]
    \midrule
\textbf{Need For Cognition} \cite{cacioppo1984efficient}  & Preference for effortful cognitive activities & “I only think as hard as I have to” \\ [0.75ex]
\textbf{Expanded Numeracy Scale} \cite{lipkus2001general}  & Objective mathematical ability & “If the chance of getting a disease is 10\% how many people would be expected to get the disease out of 100?” \\[0.75ex]
\textbf{Berlin Numeracy Test} \cite{fagerlin2007measuring}  & Objective mathematical ability & “Imagine we are throwing a loaded die (6 sides). The probability that the die shows a 6 is twice as high as the probability of each of the other numbers. On average, out of these 70 throws how many times would the die show the number 6?” \\[0.75ex]
\textbf{Subjective Numeracy Scale} \cite{cokely2012measuring} & Self-assessed mathematical ability & “How good are you at calculating a 15\% tip?” \\[0.75ex]
\textbf{Cognitive Reflection Test} \cite{frederick2005cognitive} & Tendency to override intuition & “A bat and a ball cost \$1.10 in total. The bat costs \$1.00 more than the ball. How much does the ball cost?” \\[0.75ex]
\textbf{Big Five Personality Inventory, Short Form} \cite{soto2017short} & Extraversion & “I am someone who… Is outgoing, sociable” \\[0.75ex]
 & Agreeableness  & “I am someone who… Assumes the best about others” \\[0.75ex]
 & Conscientiousness & “I am someone who… Is reliable, can always be counted on” \\[0.75ex]
 & Negative Emotionality  & “I am someone who… Tends to feel depressed, blue” \\ [0.75ex]
 & Open-Mindedness  & “I am someone who… Is original, comes up with new ideas” \\
\end{tabular}
\end{table}

Several of the personality traits that we measured are expected to be correlated since they index similar or related constructs (e.g., objective numeracy, subjective numeracy, cognitive reflection, and Need for Cognition should all be correlated somewhat \cite{liberali2012individual}). We therefore conducted a second factor analysis on users’ personality trait scores. Specifically, we conducted an Exploratory Factor Analysis (EFA) with oblique (oblimin) factor rotation and maximum likelihood factor extraction. Factors were retained if their eigenvalues were greater than or equal to 1.0 (Kaiser's criterion). 

\subsection{System Evaluation}
After completing a ``practice-run'' with a simulated system to familiarize subjects with the task, each subject was then randomly assigned to evaluate two (of the four) separate systems. Users completed system-specific training developed by each developer team. Upon completion, subjects were instructed to access their assigned system. Users tested both of their assigned  systems at their own convenience during a pre-scheduled week.

A survey was designed to evaluate users’ experience with their assigned systems. The items were originally written to capture three concepts of system utility: the usefulness, meaningfulness, and relevance of the systems’ output. Items were also included to evaluate the functionality of the system, i.e. whether the system crashed or failed to load.

The survey included 96 Likert-scale items and 12 free-response questions, including 24 items capturing usefulness, 28 items capturing meaningfulness, 24 items capturing relevance, and 20 items capturing functionality (see supplementary materials). Some items related to more than one of these concepts and were double-coded. To score the systems, protocol items were converted to numeric values. Negatively-framed questions (e.g., "I did NOT understand the model output") were reverse-coded, so that positive experiences always resulted in a positive value. Negatively- and positively-framed versions of the same item were then averaged. Items that pertained to the user's assessment of the system, and that had both forward- and reverse-coded versions, were retained for the purposes of analysis, yielding 39 composite items.  Finally, we analyzed these 39 items using an Exploratory Factor Analysis (EFA) with oblique (oblimin) factor rotation and maximum likelihood factor extraction. Factors were retained if their eigenvalues were greater than or equal to 1.0 (Kaiser's criterion).

Subjects were paid \$40/hour, for a maximum of 5 hours, for their participation unless they opted out of payment (e.g., due to employer restrictions). Subjects that did not complete their assigned tests were requested to fill out a feedback questionnaire. The study received IRB approval from IntegReview IRB. 

\subsection{Datasets}
Subject matter expert participants came from five specialty areas: conflict studies, radicalization and extremism, botany, risk assessment, and counter-terrorism.  Participants with subject matter expertise in any of these areas used specially selected datasets from their domain when evaluating their assigned systems. Participants with data science expertise, but who did not have subject matter expertise, used botany datasets to evaluate their assigned systems.  

\section{Results}

\subsection{Sample Characteristics}
86 users were recruited to participate in testing across the four datasets. Of these, 43 (50\%) users completed at least one test (32 users, 37\%, completed two tests, but we only examined users’ first tests to eliminate learning effects). Of these 43 users, 38 (88\%) also completed the full personality questionnaire. The distribution of the 38 users whose system evaluations were used in the final regressions is given in Table 2. 

\begin{table}
 \caption{Sample characteristics}
  \centering
  \begin{tabular}{l | c c c}
    \toprule
Gender & Male & 24 & 63\%  \\
 & Female & 4 & 11\%  \\
 & No response & 10 & 26\%  \\
\midrule
Race & White & 17 & 45\%  \\
 & Asian & 7 & 18\%  \\
 & Other & 4 & 11\%  \\
 & No response & 10 & 26\%  \\ 
\midrule
Ethnicity & Non-Hispanic & 27 & 71\%  \\
 & Hispanic & 7 & 3\%  \\
 & No response & 10 & 26\%  \\ 
 \midrule
Education & Obtained a graduate/professional degree & 16 & 42\%  \\
 & Graduated from a 4-year college or more & 9 & 24\%  \\
 & Attended some college but did not finish a 4-year degree & 3 & 8\%  \\ 
 & No response & 10 & 8\%  \\ 
  \midrule
  Subject Matter Expertise & Yes & 24 & 63\%  \\
 & No & 14 & 37\%  \\
  \midrule
  Current Formal Method User & Yes & 9 & 24\%  \\
 & No & 29 & 76\%  \\
  \midrule
  Current Machine Learning/AI Tool User & Yes & 20 & 43\%  \\
 & No & 18 & 47\%  \\
  \midrule
Dataset & Botany & 21 & 55\%  \\
 & Conflict Studies & 8 & 21\%  \\
 & Radicalization and Extremism & 6 & 16\%  \\ 
 & Counter-terrorism & 2 & 5\%  \\ 
 & Risk Assessment & 1 & 3\%  \\ 
   \midrule
System Tested & A & 13 & 34\%  \\
 & B & 12 & 32\%  \\
 & C & 10 & 26\%  \\ 
 & D & 2 & 5\%  \\ 
\bottomrule
  \end{tabular}
  \label{tab:table}
\end{table}
 
\subsection{Explainability, Interpretability, Data Discovery, and Ease of Visualization are Discrete Aspects of System Utility}
 
43 users completed at least one test, and therefore the complete evaluation survey. An EFA yielded 4 distinct factors, explaining 60\% of the variance in the data (see Table \ref{tab:eval_efa}). 
 
 \begin{table}
 \caption{Factor loadings for system evaluation survey data with oblimin rotation. Factors loadings > 0.4 are shown. Discovery was correlated with Interpretability, r=0.52, Explainability, r=0.37, and Visualization, r=0.50. Interpretability and Explainability were somewhat correlated r=0.33, as were Interpretability and Visualization, r=0.27. Finally,  Explainability and Visualization were weakly correlated r=0.15.}
  \begin{tabular}{p{3.2cm} c c c c }
    \toprule
\textbf{Item Short From} & Discovery (38\%) & Interpretability (12\%) & Explainability (5\%) & Visualization (5\%)\\
    \midrule
Provided.New.Insight & 0.96 &  &  & \\
Make.New.Connections & 0.95 &  &  & \\
Looked.at.Data.in.New.Way & 0.93 &  &  & \\
Explore.New.Questions & 0.89 &  &  & \\
Showed.New.Relationships & 0.88 &  &  & \\
Identify.Interesting.Hypotheses & 0.88 &  &  & \\
Think.of.Questions.in.New.Way & 0.86 &  &  & \\
Fresh.Perspective & 0.86 &  &  & \\
Helped.Think.New & 0.85 &  &  & \\
Investigate.in.New.Ways & 0.85 &  &  & \\
Identify.One.Hypothesis & 0.81 &  &  & \\
Helped.Formulate.Question & 0.79 &  &  & \\
Think.of.New.Data & -0.76 &  &  & \\
Helped.Hone.Question & 0.69 &  &  & \\
Look.for.What.I.Wanted & 0.65 &  &  & \\
Find.New.Data & 0.64 &  &  & \\
Use.New.Data & 0.64 &  &  &\\ 
New.Data.Sources & 0.58 &  &  & \\
Understand.Predictions & 0.56 &  &  & \\
Transform.to.my.Liking & 0.56 &  &  & \\
Future.Model.Performance & 0.55 &  &  & \\
Model.Trend.Performance & 0.55 &  &  & \\
Decide.What.Relevant & 0.50 &  &  &\\ 
Remove.Extra.Info & 0.48 &  &  & \\
Relevant.Trends &  &  &  & \\
Results.Make.Sense &  & 0.94 &  & \\
Explain.What.Results.Mean &  & 0.88 &  & \\
Make.Sense.of.Results &  & 0.87 &  & \\
Explain.Prediction.Reasoning &  & 0.64 & 0.54 & \\
Provided.Wanted.Info & 0.42 & 0.52 &  &\\ 
Confidence.in.Results &  & 0.51 &  & \\
Overwhelmed.with.Info & -0.44 & 0.48 &  & \\
Helped.Focus &  & 0.43 &  & \\
Aligned.With.Intuition &  & 0.42 &  & \\
Explained.Particular.Predictions &  &  & 0.51 & \\
How.Made.Predictions &  &  & 0.50 & \\
Indicated.Important.Features &  &  & 0.45 & 0.45\\
Produced.Important.Features &  &  &  & \\
Visualizations.Helped &  &  &  & 0.85\\
    \bottomrule

  \end{tabular}
  \label{tab:eval_efa}
\end{table}

These four factors measured the extent to which users were:
\begin{enumerate}
\item Able to discover new things (Discovery)
\item Able to make sense of system output (Interpretability)
\item Able to explain \textit{how} system output was generated (Explainability)
\item Able to make use of Data Visualizations (Visualization)
\end{enumerate}
As predicted, users’ abilities to explain how results were generated and to make sense of those results were statistically distinct.

\subsection{Individual Traits Predict Subject Responses}
38 users completed both the personality/individual traits questionnaire and the data-science self-report scale. These scores were used as input to a second EFA, explaining a total 53\% of the variance in the subjects' responses, yielding three distinct factors. These factors, described in Table \ref{tab:personality_efa}, can be interpreted as positivity, metacognitive monitoring and editing, and self-efficacy (i.e., one's self-assessment of one's abilities). Results for both EFAs replicated when using factors derived by varimax rotation.

\begin{table}
 \caption{Factor loadings, factor analysis of personality trait data with oblimin rotation. Factors loadings > 0.4 are shown. Positivity and Metacognition were essentially not correlated r=0.07. Positivity and Self-Efficacy were somewhat correlated r=0.22. Finally, Metacognition and Self-Efficacy were somewhat correlated r=0.37.}
  \begin{tabular}{l c c c}
    \toprule
\textbf{Item} & Positivity (20\%) & Metacognition (19\%) & Self-Efficacy (14\%) \\ 
    \midrule
Extraversion & 0.80 &  & \\
Conscientiousness & 0.68 &  & \\
Negative Emotionality & -0.57 &  & \\
Need for Cognition & 0.49 &  & \\
Agreeableness & 0.44 &  & \\
Open Mindedness &  &  & \\
Cognitive Reflection &  & 0.80 & \\
Lipkus Objective Numeracy &  & 0.79 & \\
Berlin Objective Numeracy &  & 0.78 & \\
Subjective Data Science Expertise &  &  & 0.99\\
Subjective Numeracy & 0.45 &  & 0.58\\

    \bottomrule
  
  \end{tabular}
  \label{tab:personality_efa}
\end{table}

\subsubsection{Interpretability is Associated with Metacognition, Discovery  with Familiarity}
For each dimension of system utility (Discovery, Interpretability, and Explainability, and Visualization), we conducted linear regressions using stepwise bidirectional elimination using the Akaike Information Criterion (AIC) as the selection criterion. The starting state for each regression procedure included all three personality factors as main effects, controlling for system used, dataset used, and whether the subject was a subject-matter expert, current used a formal method in their work, or currently used ML/AI tools in their work. The bidirectional stepwise algorithm progressively removed these main effects and control variables, or added statistical interactions between main effects, until a local minimum in AIC was reached(see Table \ref{tab:regression}). All results replicated when using factors derived by varimax rotation.

\begin{table}
 \caption{Outcome of stepwise regression with bidirectional elimination examining system evaluation responses as a function of personality trait factor scores and self-reported skills}
  \centering
  \begin{tabular}{ p{2cm} c c c c c c c c c c c c c c c}
    \toprule
    &  \multicolumn{3}{c}{Discovery} &  &\multicolumn{3}{c}{Interpretability} & &\multicolumn{3}{c}{Explainability} & &\multicolumn{3}{c}{Visualization}\\

DV & $\beta$ & SE & t &  & $\beta$ & SE & t &  & $\beta$ & SE & t &  & $\beta$ & SE & t\\
\midrule
System A  &  -  &  -  &  -  &    & 1.95 & 0.95 &  2.05*  &    &  -  &  -  &  - &    & -1.18 & 1.06 & -1.11\\
System B  &  -  &  -  &  -  &    & 2.14 & 0.84 & 2.56* &    &  -  &  -  &  - &    & -1.65 & 0.92 & -1.79\\
System C  &  -  &  -  &  -  &    & 1.56 & 0.82 & 1.89 &    &  -  &  -  &  - &    & -1.84 & 0.90 & -2.05*\\
System D  &  -  &  -  &  -  &    & 1.46 & 0.82 & 1.79 &    &  -  &  -  &  - &    & -2.63 & 0.89 & -2.97**\\
Risk Assessment Dataset  & -1.54 & 0.89 & -1.74 &    &  -  &  -  &  -  &    &  -  &  -  &  -  &    &  -  &  -  &  - \\
Counter-terrorism Dataset  & -1.47 & 0.64 & -2.32* &    &  -  &  -  &  -  &    &  -  &  -  &  -  &    &  -  &  -  &  - \\
Radicalization and Extremism Dataset  & -0.54 & 0.45 & -1.20 &    &  -  &  -  &  -  &    &  -  &  -  &  -  &    &  -  &  -  &  - \\
Botany Dataset  & -0.82 & 0.39 & -2.13* &    &  -  &  -  &  -  &    &  -  &  -  &  -  &    &  -  &  -  &  -\\ 
Use Formal Method?  & 0.93 & 0.33 & 2.82** &    &  -  &  -  &  -  &    & 0.64 & 0.36 & 1.76 &    &  -  &  -  &  - \\
Use ML/AI?  & 0.49 & 0.32 & 1.52 &    & 0.54 & 0.29 & 1.87 &    &  -  &  -  &  - &    & 0.44 & 0.29 & 1.52\\
Subject Matter Expert?  &  -  &  -  &  -  &    &  -  &  -  &  -  &    &  -  &  -  &  - &    &  -  &  -  &  -\\
Positivity  & 0.28 & 0.16 & 1.73 &    &  -  &  -  &  -  &    &  -  &  -  &  - &    & 0.31 & 0.17 & 1.85\\
Metacognition  &  -  &  -  &  -  &    & 0.53 & 0.23 & 2.35* &    &  -  &  -  &  - &    &  -  &  -  &  -\\
Self-Efficacy  &  -  &  -  &  -  &    &  -  &  -  &  -  &    &  -  &  -  &  - &    &  -  &  -  &  -\\
Intercept  & 0.13 & 0.29 & 0.44 &    & -2.12 & 0.78 & -2.73* &    & -0.20 & 0.18 & -1.14 &    & 1.68 & 0.85 & 1.97\\
    \bottomrule
  \end{tabular}
  \label{tab:regression}
\end{table}

\section{Discussion}
Our results provide support for Fuzzy-Trace Theory's predictions. Specifically, we find that interpretability and explainability are statistically distinct, as demonstrated by the fact that they loaded separately on a factor analysis. They are nevertheless somewhat correlated. This indicates that there are likely to be some subjects that are able to derive meaning from an understanding of how a system operates, but that the majority of subjects could not. 

According to Fuzzy-Trace Theory, individual differences mediate the extent to which subjects rely on different mental representations. Specifically, factors associated with metacognitive monitoring and editing -- such as numeracy and Need for Cognition -- should predict reliance on increasingly precise mental representations. Our data show that a factor associated with these metacognitive traits is indeed associated with increased interpretability assessment, even after controlling for the specific system that generated model output. 

Additionally, we found that a subject's familiarity with formal methods was associated with discovery, controlling for dataset. In effect, these results suggest that the ability to generate a new result using an automated machine-learning system may be primarily driven by experience with that system. Notably, no significant predictors were associated with changes in explainability. Furthermore, we did not detect any changes in discovery or explainability between systems. Importantly, there was a difference in discovery between datasets, with the botany and counter-terrorism datasets both more difficult to use for discovery purposes. 

Beyond these factors, our results suggest that in order to understand that result and apply it to a real-world problem, it must be interpreted. The ability to generate these interpretations, in turn, may depend on individual personality traits, such as those associated with mathematical ability and gist processing. Furthermore, the fact that interpretability is distinct from visualization indicates that communicating that meaning of a machine learning model's results is not simply just a matter of improved visualization. 

Notably, the strongest predictor of whether a subject was able to generate a meaningful interpretation was whether they used ``System A''. This means that it is indeed possible to design systems that promote interpretability, and that these systems may not be the same as those that promote explainability. 

Overall, our results suggest that interpretability depends on mental representation. Any algorithms that strives for interpretability must take mental representation into account, highlighting the need for design work to first extract these representations. Since users differ in both their background knowledge and in their individual traits, interpretable systems may have to tailor their outputs to the specific needs of their users. Importantly, these features are not entirely random; rather, our data seem to indicate that they depend on well-known and characterized predispositions to metacognitive monitoring and editing, and less so on personality traits that vary significantly between subjects (e.g., the ``Big Five'').

\subsection{Limitations}
Due to the small size of our sample, it is likely that there are confounding factors at play. For example, the botany dataset contained the most subjects because all subjects without subject-matter expertise were assigned to it. Additionally, subjects were not evenly balanced between datasets, systems, and prior use of formal methods and ML/AI, in large part because not all recruited users completed testing. Future work must therefore more explicitly control for these factors. Finally, the small sample size leaves us unable to examine several theorized interactions, such as between subject-matter expertise and metacognition.

This work is also limited by the fact that data science expertise was self-reported. Indeed, our factor analysis results show that data science expertise loaded with subjective numeracy rather than with Cognitive Reflection and objective numeracy, meaning that it may be more of a measure of self-efficacy than of actual skill. Creating a reliable, theoretically motivated measure of data science expertise is an exciting opportunity for future work.

\subsection{Conclusions}
In conclusion, our analysis offers preliminary evidence for the distinct contributions of interpretability and explainability to system utility. Per Fuzzy-Trace Theory, interpretability should be associated with less precise, yet productive, gist processing, whereas explainability may be more associated with the ability to debug systems, but not necessarily the ability to apply their output to solve real-world problems. Both skill sets are necessary, motivating the need for more research into systems, and users, that can translate between these different mental representations when machine learning tools are used in high-stakes settings.

\bibliographystyle{unsrt}  
\bibliography{references}  

\begin{landscape}
\begin{longtable}{p{6cm} p{1.5cm} p{11cm} p{2cm}}
    \multicolumn{2}{c}{\tablename\ \thetable\ -- \textit{Continued from previous page}}\\[12pt]
    \hline
    \endhead
    \hline
    \multicolumn{2}{r}{\tablename\ \thetable\ -- \textit{Continued on next page}} \\
    \endfoot
    \caption{System evaluation item details and responses}
\\     \hline
     \textbf{Item Short Form} & \textbf{Code} & \textbf{Item Full Text} & \textbf{Mean (SD)} \\ [0.75ex]
     \hline
     \endfirsthead
 \endlastfoot
 Explain.What.Results.Mean & (+) & I can explain what the system's results mean & 0.15 (1.93) \\ [0.75ex]
 & (-) & I CANNOT explain what these results mean & 0.02 (1.99) \\ [0.75ex]
    \hline
Make.Sense.of.Results & (+) & I can make sense of what the system's results are saying & 0.44 (1.75) \\ [0.75ex]
 & (-) & I can NOT make sense of what the system's results are saying & -0.13 (1.94) \\ [0.75ex]
    \hline
Results.Make.Sense & (+) & The system's results make sense to me & 0.29 (1.87) \\ [0.75ex]
 & (-) & The system's results do NOT make sense to me & -0.05 (1.99) \\ [0.75ex]
    \hline
Explain.Prediction.Reasoning & (+) & I can explain the reasoning behind the predictions the model made & -0.02 (1.68) \\ [0.75ex]
 & (-) & I CANNOT explain the reasoning behind the predictions the model made & 0.02 (1.92) \\ [0.75ex]
    \hline
Provided.New.Insight & (+) & The system provided me with new insight & 0.07 (1.59) \\ [0.75ex]
 & (-) & The system did NOT provide me with new insight & -0.12 (1.49) \\ [0.75ex]
    \hline
Aligned.With.Intuition & (+) & The system's output aligned with my intuition & 0.46 (1.32) \\ [0.75ex]
 & (-) & The system's output did NOT align with my intuition & -0.54 (1.4) \\ [0.75ex]
    \hline
Investigate.in.New.Ways & (+) & The system allowed me to investigate the data in new ways & 0.54 (1.55) \\ [0.75ex]
 & (-) & The system did NOT allow me to investigate the data in new ways & -0.22 (1.49) \\ [0.75ex]
    \hline
Transform.to.my.Liking & (+) & The system allowed me to transform the data to my liking & -0.1 (1.55) \\ [0.75ex]
 & (-) & The system did NOT allow me to transform the data to my liking & 0.29 (1.49) \\ [0.75ex]
    \hline
Confidence.in.Results & (+) & I have confidence in the system's results & 0.66 (1.7) \\ [0.75ex]
 & (-) & I do NOT have confidence in the system's results & -0.49 (1.53) \\ [0.75ex]
    \hline
Overwhelmed.with.Info & (-) & The system gave me an overwhelming amount of information & -0.49 (1.68) \\ [0.75ex]
 & (+) & The system did NOT give me an overwhelming amount of information & 0.22 (1.78) \\ [0.75ex]
    \hline
Understand.Predictions & (+) & The model allowed me to understand individual predictions based on these data & 0.05 (1.7) \\ [0.75ex]
 & (-) & The model did NOT allow me to understand individual predictions based on these data & 0.02 (1.59) \\ [0.75ex]
    \hline
Remove.Extra.Info  & (+) & The system helped me to remove extraneous information & -0.07 (1.59) \\ [0.75ex]
 & (-) & The system did NOT help me remove extraneous information & 0.12 (1.66) \\ [0.75ex]
    \hline
Produced.Important.Features  & (+) & The model produced features which I deemed important & 0.37 (1.67) \\ [0.75ex]
 & (-) & The model did NOT produce features which I deem important & -0.24 (1.58) \\ [0.75ex]
    \hline
Helped.Focus & (+) & The system helped me focus on what mattered to perform the task & 0.49 (1.5) \\ [0.75ex]
 & (-) & The system did NOT help me focus on what mattered to perform the task & -0.27 (1.6) \\ [0.75ex]
    \hline
Indicated.Important.Features  & (+) & The system indicated which features were important & -0.02 (1.74) \\ [0.75ex]
 & (-) & The system did NOT indicate which features were important & -0.02 (1.75) \\ [0.75ex]
    \hline
How.Made.Predictions & (+) & The system allowed me to see how it made predictions & -0.05 (1.58) \\ [0.75ex]
 & (-) & The system did NOT allow me to see how it made predictions & -0.15 (1.51) \\ [0.75ex]
    \hline
Explained.Particular.Predictions & (+) & The system explained to me how I got a particular prediction & -0.41 (1.47) \\ [0.75ex]
 & (-) & The system did NOT explain to me how I got a particular prediction & 0.46 (1.55) \\ [0.75ex]
    \hline
Helped.Think.New & (+) & The system helped me to think of something new & 0.2 (1.55) \\ [0.75ex]
 & (-) & The system did NOT help me to think of something new & -0.22 (1.67) \\ [0.75ex]
    \hline
Helped.Hone.Question & (+) & The system helped me to hone the question(s) that I was asking & 0.12 (1.55) \\ [0.75ex]
 & (-) & The system did NOT help me to hone the question(s) that I was asking & -0.05 (1.6) \\ [0.75ex]
    \hline
Helped.Formulate.Question & (+) & The system helped me to better formulate my question(s) & 0.34 (1.56) \\ [0.75ex]
 & (-) & The system did NOT help me to better formulate my question(s) & 0.07 (1.65) \\ [0.75ex]
    \hline
Explore.New.Questions & (+) & The system allowed me to explore new questions as I discovered them & 0.29 (1.55) \\ [0.75ex]
 & (-) & The system did NOT allow me to explore new questions as I discovered them & -0.37 (1.5) \\ [0.75ex]
    \hline
Identify.One.Hypothesis & (+) & The system helped me to identify at least one interesting hypothesis to test & 0.76 (1.43) \\ [0.75ex]
 & (-) & The system did NOT help me to identify at least one interesting hypothesis to test & -0.59 (1.61) \\ [0.75ex]
    \hline
Identify.Interesting.Hypotheses & (+) & The system helped me to identify many new, interesting hypotheses & 0.05 (1.41) \\ [0.75ex]
 & (-) & The system did NOT help me to identify very many new, interesting hypotheses & 0.02 (1.65) \\ [0.75ex]
    \hline
Make.New.Connections  & (+) & The system helped me make connections I had not previously thought of & 0.32 (1.51) \\ [0.75ex]
 & (-) & The system did NOT help me make connections I had not previously thought of & -0.24 (1.58) \\ [0.75ex]
    \hline
Showed.New.Relationships & (+) & The system showed me new and interesting relationships in the data & 0.49 (1.45) \\ [0.75ex]
 & (-) & The system did NOT show me new and interesting relationships in the data & -0.37 (1.53) \\ [0.75ex]
    \hline
Look.for.What.I.Wanted & (+) & The system allowed me to look for what I wanted in the data & 0.37 (1.64) \\ [0.75ex]
 & (-) & The system did NOT allow me to look for what I wanted in the data & -0.07 (1.54) \\ [0.75ex]
    \hline
Provided.Wanted.Info & (+) & The system was able to provide me with information that I wanted & 0.29 (1.65) \\ [0.75ex]
 & (-) & The system was NOT able to provide me with information that I wanted & 0 (1.69) \\ [0.75ex]
    \hline
Looked.at.Data.in.New.Way & (+) & The system helped me to look at my data in a new way & 0.44 (1.53) \\ [0.75ex]
 & (-) & The system did NOT help me to look at my data in a new way & -0.17 (1.5) \\ [0.75ex]
    \hline
Think.of.Questions.in.New.Way  & (+) & The system helped me to think of the question(s) I work on in a new way & 0.41 (1.58) \\ [0.75ex]
 & (-) & The system did NOT help me to think of the question(s) I work on in a new way & -0.1 (1.61) \\ [0.75ex]
    \hline
Think.of.New.Data & (+) & The system helped me think of new data sources to use in my work & 0.29 (1.54) \\ [0.75ex]
 & (-) & The system did NOT help me think of new data sources to use in my work & 0.15 (1.57) \\ [0.75ex]
    \hline
Decide.What.Relevant & (+) & The system helped me to decide what information was relevant & 0.44 (1.64) \\ [0.75ex]
 & (-) & The system did NOT help me to decide what information was relevant & 0.1 (1.83) \\ [0.75ex]
    \hline
Use.New.Data & (+) & The system made effective use of new sources of data & 0.07 (1.35) \\ [0.75ex]
 & (-) & The system did NOT make effective use of new sources of data & -0.22 (1.37) \\ [0.75ex]
    \hline
New.Data.Sources & (+) & The system helped me to find new sources of data easily & 0.15 (1.41) \\ [0.75ex]
 & (-) & The system did NOT help me to find new sources of data easily & 0.02 (1.49) \\ [0.75ex]
    \hline
Find.New.Data & (+) & The system made it easy to find new sources of relevant data & -0.07 (1.51) \\ [0.75ex]
 & (-) & The system did NOT make it easy to find new sources of relevant data & 0.07 (1.6) \\ [0.75ex]
    \hline
Fresh.Perspective & (+) & The system helped provide a fresh perspective & 0.24 (1.7) \\ [0.75ex]
 & (-) & The system did NOT help provide a fresh perspective & -0.22 (1.57) \\ [0.75ex]
    \hline
Visualizations.Helped & (+) & The system's visualizations helped me to understand the task at hand & 0.68 (1.62) \\ [0.75ex]
 & (-) & The system's visualizations did NOT help me to understand the task at hand & -0.15 (1.78) \\ [0.75ex]
    \hline
Model.Trend.Performance & (+) & The system allowed me to investigate how well the model performed in finding relationships in prior data & 0.27 (1.5) \\ [0.75ex]
 & (-) & The system did NOT allow me to investigate how well the model performed in finding relationships in prior data & 0.24 (1.5) \\ [0.75ex]
    \hline
Future.Model.Performance & (+) & The system helped me to use information about relationships based on prior data to make predictions about future model performance & 0.27 (1.48) \\ [0.75ex]
 & (-) & The system did NOT help me to use information about relationships based on prior data to make predictions about future model performance & -0.37 (1.43) \\ [0.75ex]
    \hline
Decide.What.Relevant & (+) & The system helped me understand which prior data and trends were relevant & 0.32 (1.65) \\ [0.75ex]
 & (-) & The system did NOT help me understand which prior data and trends were relevant & -0.22 (1.49) \\ [0.75ex]
 \hline

\end{longtable}
\end{landscape}

\end{document}